\documentclass[english]{revtex4-1}
\usepackage[T1]{fontenc}
\usepackage[latin9]{inputenc}
\usepackage{amsmath}
\usepackage{amssymb}

\makeatletter
\usepackage{babel}

\makeatother

\usepackage{babel}
\begin{document}

\title{The minimum width in relativistic quantum mechanics}

\author{Scott E. Hoffmann}

\address{School of Mathematics and Physics,~~\\
 The University of Queensland,~~\\
 Brisbane, QLD 4072~~\\
 Australia}
\email{scott.hoffmann@uqconnect.edu.au}

\selectlanguage{english}%
\begin{abstract}
We challenge the widespread belief, originated by Newton and Wigner
(Rev. Mod. Phys, \textbf{21}, 400 (1949)) that the incorporation of
special relativity into quantum mechanics implies that a massive particle
cannot be localized within an arbitrarily small spatial extent, that
there is a minimum width approximately equal to the Compton wavelength.
Our argument is in four parts. First, the scalar function used by
Newton and Wigner as a measure of localization is not a position probability
amplitude. The correct relativistic position probability amplitude
becomes a delta function for a state vector localized according to
the criteria of Newton and Wigner. Second, the possibility of Lorentz
contraction as observed from a boosted frame means that the wavepacket
width in the boost direction can take arbitrarily small values. Third,
we refer to the work of Almeida and Jabs (Am. J. Phys. \textbf{52},
921 (1984)) who show that the long time wavepacket spreading rate
for relativistic position probability amplitudes is always less than
the speed of light no matter how small the initial width of the wavepacket.
Lastly, we show that it is a simple matter to construct scalar amplitudes
with spatial widths smaller than the supposed minimum.
\end{abstract}
\maketitle

\section{Introduction}

The aim of this paper is to challenge a widely held belief, expressed
first by Newton and Wigner \cite{Newton1949} and repeated elsewhere
\cite{Horwitz2015,Greiner2000,Sakurai1967,Feshbach1958}. This belief
is that the incorporation of special relativity into quantum mechanics
implies that a massive, spinless, particle localized according to
the Newton-Wigner criteria has a nonzero spatial extent, a minimum
width approximately equal to its Compton wavelength. In support of
this conclusion, it has been argued \cite{Greiner2000,Itzykson1980}
that attempting to localize an electron to within a Compton wavelength
could use energy more than twice the rest energy of the electron,
so that electron-positron pair creation would result. We will deal
with this argument, but note first that it is puzzling why a theory
to describe a \textit{free} particle would have anything to say about
pair creation.

The argument against this position presented here has four parts.
First, the quantity used by Newton and Wigner to characterize the
spatial extent of the localized particle is not a position probability
amplitude. Its modulus-squared does not have the properties required
of a position probability density. In contrast, such a position probability
amplitude \textit{can} be defined in the relativistic context \cite{Hoffmann2018d,Rosenstein1985,Fong1968,Foldy1956}
and its modulus-squared satisfies all the conditions required of a
position probability density. The second argument against a minimum
width is that such a statement is not consistent with the possibility
of Lorentz contraction of the wavepacket. In Section III we demonstrate
the Lorentz contraction of a position probability amplitude and compare
with the case of a scalar amplitude. The result is that a particle
can be localized in an arbitrarily small volume as observed from a
boosted frame, with no pair creation taking place. Thirdly, we refer
to the work of Almeida and Jabs \cite{Almeida1984}, who calculated
the relativistic spreading of position probability amplitudes for
initial extents small compared to the Compton wavelength. They found
that the initial wavepacket could have an arbitrarily small spatial
extent without causing the long time spreading rate to exceed the
speed of light. Lastly, it is a simple matter to construct well-behaved
scalar amplitudes with spatial widths smaller than the Compton wavelength.

The organization of this paper is as follows. In Section II we review
the properties of relativistic probability amplitudes and consider
the results of Newton and Wigner. In Section III we calculate the
Lorentz contraction of a position probability amplitude and of a scalar
amplitude. In Section IV we review the results of Almeida and Jabs
on the spreading of relativistic probability amplitudes. In Section
V we construct scalar amplitudes with spatial widths less than the
Compton wavelength. Conclusions follow.

Throughout this paper, we use Heaviside-Lorentz units, in which $\hbar=c=\epsilon_{0}=\mu_{0}=1$.

\section{Position probability amplitudes and the Newton-Wigner result}

The wavefunctions that we call relativistic probability amplitudes
have been studied before \cite{Hoffmann2018d,Rosenstein1985,Fong1968,Foldy1956}.
They are called Newton-Wigner-Foldy wavefunctions by Rosenstein \cite{Rosenstein1985}.
Here we briefly review their definition. For a fuller account, including
their transformation properties, we refer the reader to a paper in
preparation by this author \cite{Hoffmann2018d}.

The set of basis vectors we use for a free, massive, spinless particle
is $\{|\,p\,\rangle:\boldsymbol{p}\in\mathbb{R}^{3},p^{0}=\omega=\sqrt{\boldsymbol{p}^{2}+m^{2}}\},$
eigenvectors of the four-momentum operator $\hat{P}^{\mu}$ with eigenvalues
$p^{\mu}$, for components with $\mu=0,1,2,3.$ Only positive energies
are considered. These basis vectors carry the unitary, irreducible
representations of the Poincaré group. We choose to use the invariant
normalization,
\begin{equation}
\langle\,p_{1}\,|\,p_{2}\,\rangle=\omega_{1}\delta^{3}(\boldsymbol{p}_{1}-\boldsymbol{p}_{2}),\label{eq:2.1}
\end{equation}
with $\omega_{1}=\sqrt{\boldsymbol{p}_{1}^{2}+m^{2}},$ as this makes
the transformation properties of the basis vectors take the simplest
forms. A normalized state vector can be written
\begin{equation}
|\,\psi\,\rangle=\int\frac{d^{3}p}{\sqrt{\omega}}\,|\,p\,\rangle\Psi(p).\label{eq:2.2}
\end{equation}
In this form, $\Psi(p)$ can be interpreted as a momentum probability
amplitude, since its normalization condition is
\begin{equation}
\int d^{3}p\,|\Psi(p)|^{2}=1\label{eq:2.3}
\end{equation}
and the expectation of four-momentum is
\begin{equation}
\langle\,\psi\,|\,\hat{P}^{\mu}\,|\,\psi\,\rangle=\int d^{3}p\,|\Psi(p)|^{2}\,p^{\mu}.\label{eq:2.4}
\end{equation}

The Newton-Wigner (improper) state vector for a particle localized
at the origin at $t=0$ is \cite{Newton1949}
\begin{equation}
|\,0,\boldsymbol{0}\,\rangle=\int\frac{d^{3}p}{\sqrt{\omega}}\,|\,p\,\rangle\frac{1}{(2\pi)^{\frac{3}{2}}}.\label{eq:2.5}
\end{equation}
A spacetime translation then gives
\begin{equation}
|\,x\,\rangle=\int\frac{d^{3}p}{\sqrt{\omega}}\,|\,p\,\rangle\frac{e^{ip\cdot x}}{(2\pi)^{\frac{3}{2}}}.\label{eq:2.6}
\end{equation}
These obey the localization condition at equal times
\begin{equation}
\langle\,t,\boldsymbol{x}\,|\,t,\boldsymbol{y}\,\rangle=\delta^{3}(\boldsymbol{x}-\boldsymbol{y}).\label{eq:2.7}
\end{equation}

Then we form the amplitudes, functions of position and time ($x^{\mu}=(t,\boldsymbol{x})^{\mu}$),
\begin{equation}
\psi(x)=\langle\,x\,|\,\psi\,\rangle=\int\frac{d^{3}p}{(2\pi)^{\frac{3}{2}}}e^{-ip\cdot x}\,\Psi(p),\label{eq:2.8}
\end{equation}
as in the nonrelativistic theory. The position operator at $t=0,$
\begin{equation}
\hat{\boldsymbol{x}}(0)=\int d^{3}x\,|\,0,\boldsymbol{x}\,\rangle\,\boldsymbol{x}\,\langle\,0,\boldsymbol{x}\,|,\label{eq:2.9}
\end{equation}
can be represented by
\begin{equation}
\hat{\boldsymbol{x}}(0)=i\frac{\partial}{\partial\boldsymbol{p}}\label{eq:2.10}
\end{equation}
acting on wavefunctions $\Psi(p),$ such that
\begin{equation}
\langle\,\psi\,|\,\hat{\boldsymbol{x}}(0)\,|\,\psi\,\rangle=\int d^{3}p\,\Psi^{*}(p)\,i\frac{\partial}{\partial\boldsymbol{p}}\Psi(p)=\int d^{3}x\,|\,\psi(0,\boldsymbol{x})|^{2}\,\boldsymbol{x}.\label{eq:2.11}
\end{equation}
The result at arbitrary times is
\begin{equation}
\langle\,\psi\,|\,\hat{\boldsymbol{x}}(t)\,|\,\psi\,\rangle=\int d^{3}p\,\Psi^{*}(p)\,(i\frac{\partial}{\partial\boldsymbol{p}}+\frac{\boldsymbol{p}}{\omega}t)\,\Psi(p)=\int d^{3}x\,|\,\psi(t,\boldsymbol{x})|^{2}\,\boldsymbol{x}.\label{eq:2.12}
\end{equation}
So this amplitude, $\psi(x),$ can be interpreted as a position probability
amplitude. From the Parseval theorem, it is normalized at all times:
\begin{equation}
\int d^{3}x\,|\,\psi(t,x)|^{2}=\int d^{3}p\,|\Psi(p)|^{2}=1.\label{eq:2.13}
\end{equation}

Newton and Wigner \cite{Newton1949} claim that the position operator
at $t=0$ is
\begin{equation}
\hat{\boldsymbol{x}}_{\mathrm{NW}}(0)=i\frac{\partial}{\partial\boldsymbol{p}}-i\frac{\boldsymbol{p}}{2\omega^{2}},\label{eq:2.14}
\end{equation}
but theirs is defined to act on scalar amplitudes $\Phi(p)=\sqrt{\omega}\,\Psi(p),$
not momentum probability amplitudes $\Psi(p).$ It is, in fact, the
same operator when acting on the same wavefunctions. The apparent
difference is resolved in the identity
\begin{equation}
\int\frac{d^{3}p}{\omega}\,\Phi^{(1)*}(p)\{i\frac{\partial}{\partial\boldsymbol{p}}-i\frac{\boldsymbol{p}}{2\omega^{2}}\}\Phi^{(2)}(p)=\int d^{3}p\,\Psi^{(1)*}(p)\{i\frac{\partial}{\partial\boldsymbol{p}}\}\Psi^{(2)}(p).\label{eq:2.15}
\end{equation}
This fact was noted in \cite{Fong1968,Hoffmann2018d}.

Both $\Psi(p)$ and $\psi(x)$ have well-defined transformations under
translations, rotations, boosts, space inversion and time reversal.
These are given in the general massive case for particles with spin
in \cite{Hoffmann2018d}. We note that the boost transformation of
$\psi(x)$ is nonlocal, as it must be to preserve total probability.
As stated in \cite{Hoffmann2018d}, special relativity does not require
that all physically relevant quantities transform as scalars, four-vectors
and tensors. It merely requires that all such transformation properties
be well-defined and depend only on the translation, rotation and boost
parameters. In quantum mechanics, these transformations must be unitary
for Lorentz transformations and space inversion, antiunitary for time
reversal.

Newton and Wigner use a function that transforms as a scalar function
as their measure of localization:
\begin{equation}
\varphi(x)=\int\frac{d^{3}p}{(2\pi)^{\frac{3}{2}}\omega}e^{-ip\cdot x}\,\sqrt{\omega}\,\Psi(p).\label{eq:2.16}
\end{equation}
Since the momentum components of this quantity are multiplied by $1/\sqrt{\omega}$
compared to those of $\psi(x),$ $\varphi(x)$ is smeared in position
space compared to $\psi(x).$ Hence, the position probability amplitude
for the particle localized at the origin ($\Psi_{0}(p)=1/(2\pi)^{\frac{3}{2}}$)
is
\begin{equation}
\psi_{0}(0,\boldsymbol{x})=\delta^{3}(\boldsymbol{x}),\label{eq:2.17}
\end{equation}
while the scalar function is
\begin{equation}
\varphi_{0}(0,\boldsymbol{x})=\int\frac{d^{3}p}{(2\pi)^{\frac{3}{2}}\sqrt{\omega}}\,\frac{e^{i\boldsymbol{p}\cdot\boldsymbol{x}}}{(2\pi)^{\frac{3}{2}}}\propto(\frac{m}{r})^{\frac{5}{4}}H_{\frac{5}{4}}^{(1)}(imr).\label{eq:2.18}
\end{equation}
This latter has the asymptotic form $\exp(-mr),$ so the relevant
spatial width is the Compton wavelength, $\lambda_{\mathrm{C}}=1/m$.
In contrast, the position probability amplitude, a delta function
with vanishing width, measures the localization correctly. So we see
the origin of the misconception: analysis of the spatial distribution
of a particle must use the relativistic probability amplitudes.

\section{Lorentz contraction}

We consider the normalized state vector
\begin{equation}
|\,\psi\,\rangle=\int\frac{d^{3}p}{\sqrt{\omega}}\,|\,p\,\rangle\Psi(p)=\int\frac{d^{3}p}{\sqrt{\omega}}\,|\,p\,\rangle\frac{e^{-|\boldsymbol{p}|^{2}/4\sigma_{p}^{2}}}{(2\pi\sigma_{p}^{2})^{\frac{3}{4}}},\label{eq:13}
\end{equation}
representing a spinless particle with vanishing average momentum.
The position probability amplitude at $t=0$ is
\begin{equation}
\psi(0,\boldsymbol{x})=\frac{e^{-|\boldsymbol{x}|^{2}/4\sigma_{x}^{2}}}{(2\pi\sigma_{x}^{2})^{\frac{3}{4}}},\label{eq:14}
\end{equation}
with $\sigma_{x}\sigma_{p}=1/2.$ So the wavepacket is minimal and
localized around the origin with a spatial width $\sigma_{x}$ in
all directions at this time.

Using the transformation results from \cite{Hoffmann2018d}, a boost
by velocity $\boldsymbol{\beta}_{0}$ produces the momentum wavefunction
\begin{equation}
\Psi^{\prime}(p)=\sqrt{\gamma_{0}(1-\boldsymbol{\beta}_{0}\cdot\boldsymbol{\beta})}\,\Psi(\Lambda^{-1}p),\label{eq:15}
\end{equation}
with $\gamma_{0}=1/\sqrt{1-\beta_{0}^{2}}$ and $\boldsymbol{\beta}=\boldsymbol{p}/\omega.$
The boosted position wavefunction at $t=0$ is then
\begin{equation}
\psi^{\prime}(0,x)=\int\frac{d^{3}p}{(2\pi)^{\frac{3}{2}}}e^{ip\cdot x}\,\sqrt{\gamma_{0}(1-\boldsymbol{\beta}_{0}\cdot\boldsymbol{\beta})}\,\Psi(\Lambda^{-1}p).\label{eq:15.1}
\end{equation}

With $p^{\prime}=\Lambda^{-1}p,$ we have
\begin{equation}
\boldsymbol{p}^{\prime}=\boldsymbol{p}_{\perp}+\gamma_{0}(\boldsymbol{p}_{\parallel}-\boldsymbol{\beta}_{0}\omega),\label{eq:16}
\end{equation}
where we have separated the momentum into parts parallel ($\boldsymbol{p}_{\parallel}$)
and perpendicular ($\boldsymbol{p}_{\perp}$) to the boost velocity.
So the exponent contains a factor
\begin{align}
|\boldsymbol{p}^{\prime}|^{2} & =|\boldsymbol{p}_{\perp}|^{2}+\gamma_{0}^{2}|\boldsymbol{p}_{\parallel}-\boldsymbol{\beta}_{0}\omega|^{2}.\label{eq:17}
\end{align}
We find that this vanishes where
\begin{equation}
\boldsymbol{p}_{\perp}=0\quad\mathrm{and}\quad\boldsymbol{p}_{\parallel}=m\gamma_{0}\boldsymbol{\beta}_{0},\label{eq:18}
\end{equation}
so the modulus-squared of the wavefunction will have its peak there.

We expand Eq. (\ref{eq:17}) in powers of $\boldsymbol{p}_{\perp}$
and $\boldsymbol{p}_{\parallel}-m\gamma_{0}\boldsymbol{\beta}_{0},$
to find
\begin{align}
|\boldsymbol{p}^{\prime}|^{2} & \cong|\boldsymbol{p}_{\perp}|^{2}+\frac{1}{\gamma_{0}^{2}}|\boldsymbol{p}_{\parallel}-m\gamma_{0}\boldsymbol{\beta}_{0}|^{2}.\label{eq:19}
\end{align}
The correction terms are of third order in the expansion quantities.
Note that the width in momentum in the boost direction will be enlarged
by the gamma factor. We choose the momentum width, $\sigma_{p},$
so that the boosted wavefunction will be narrow in momentum:
\begin{equation}
\frac{\gamma_{0}\sigma_{p}}{m\gamma_{0}\beta_{0}}=\frac{\sigma_{p}}{m\beta_{0}}\ll1.\label{eq:20}
\end{equation}
Then the third order terms we neglected in Eq. (\ref{eq:19}) will
produce higher powers of this small ratio, so are justifiably negligible.

Also we take $\boldsymbol{\beta}\rightarrow\boldsymbol{\beta}_{0},$
its peak value, in the slowly varying factor in Eq. (\ref{eq:15}),
giving
\begin{equation}
\sqrt{\gamma_{0}(1-\boldsymbol{\beta}_{0}\cdot\boldsymbol{\beta})}\rightarrow\frac{1}{\sqrt{\gamma_{0}}}.\label{eq:21}
\end{equation}

The position wavefunction at $t=0$ is then, defining $\boldsymbol{x}_{\parallel}$
and $\boldsymbol{x}_{\perp}$ as components of the position parallel
and perpendicular to $\boldsymbol{\beta}_{0},$ respectively,
\begin{equation}
\psi^{\prime}(0,\boldsymbol{x})\cong\int\frac{d^{3}p}{(2\pi)^{\frac{3}{2}}}e^{i(\boldsymbol{p}_{\perp}\cdot\boldsymbol{x}_{\perp}+\boldsymbol{p}_{\parallel}\cdot\boldsymbol{x}_{\parallel})}\,\frac{1}{\gamma_{0}^{\frac{1}{2}}}\,\frac{e^{-|\boldsymbol{p}_{\perp}|^{2}/4\sigma_{p}^{2}}e^{-|\boldsymbol{p}_{\parallel}-m\gamma_{0}\boldsymbol{\beta}_{0}|^{2}/4\gamma_{0}^{2}\sigma_{p}^{2}}}{(2\pi\sigma_{p}^{2})^{\frac{3}{4}}}.\label{eq:22}
\end{equation}
We evaluate the integrals using \cite{Gradsteyn1980} (their formula
3.323.2) in the form
\begin{equation}
\int_{-\infty}^{\infty}dz\,e^{-z^{2}/\sigma^{2}}e^{i\xi\,z/\sigma}=(\pi\sigma^{2})^{\frac{1}{2}}e^{-\xi^{2}/4}.\label{eq:22.1}
\end{equation}
This gives
\begin{equation}
\psi^{\prime}(0,\boldsymbol{x})\cong e^{im\gamma_{0}\boldsymbol{\beta}_{0}\cdot\boldsymbol{x}_{\parallel}}\frac{e^{-|\boldsymbol{x}_{\perp}|^{2}/4\sigma_{x}^{2}}}{(2\pi\sigma_{x}^{2})^{\frac{1}{2}}}\frac{e^{-\gamma_{0}^{2}|\boldsymbol{x}_{\parallel}|^{2}/4\sigma_{x}^{2}}}{(2\pi\sigma_{x}^{2}/\gamma_{0}^{2})}.\label{eq:23}
\end{equation}
Thus we see the Lorentz contraction of the spatial width in the boost
direction,
\begin{equation}
\sigma_{x\parallel}\rightarrow\frac{\sigma_{x}}{\gamma_{0}}.\label{eq:24}
\end{equation}
Since $\gamma_{0}$ can take arbitrarily large values, the particle
can be localized in an arbitrarily small volume.

If high energy photons were used to localize an electron in a small
volume, electron-positron pair creation would indeed occur. The method
of strong localization considered here also requires a large amount
of energy, but only from the propellant used to boost the observer.
Both observers would confirm that no pair creation was taking place.

We perform a very similar calculation to find the boost behaviour
of the scalar amplitude for the same choice of initial state vector
\begin{equation}
\varphi^{\prime}(0,\boldsymbol{x})=\int\frac{d^{3}p}{(2\pi)^{\frac{3}{2}}\omega}e^{i\boldsymbol{p}\cdot\boldsymbol{x}}\,\sqrt{\gamma_{0}(1-\boldsymbol{\beta}_{0}\cdot\boldsymbol{\beta})}\,\Psi(\Lambda^{-1}p).\label{eq:25-1}
\end{equation}
We replace the slowly varying factor $1/\omega$ by its value at the
momentum wavefunction peak,
\begin{equation}
\frac{1}{\omega}\rightarrow\frac{1}{\sqrt{m^{2}+m^{2}\gamma_{0}^{2}\beta_{0}^{2}}}=\frac{1}{m\gamma_{0}}.\label{eq:25-2}
\end{equation}
So the result is (with the correct dimension)
\begin{equation}
\varphi^{\prime}(0,\boldsymbol{x})\cong\frac{1}{m\gamma_{0}}e^{im\gamma_{0}\boldsymbol{\beta}_{0}\cdot\boldsymbol{x}_{\parallel}}\frac{e^{-|\boldsymbol{x}_{\perp}|^{2}/4\sigma_{x}^{2}}}{(2\pi\sigma_{x}^{2})^{\frac{1}{2}}}\frac{e^{-\gamma_{0}^{2}|\boldsymbol{x}_{\parallel}|^{2}/4\sigma_{x}^{2}}}{(2\pi\sigma_{x}^{2}/\gamma_{0}^{2})}.\label{eq:25-3}
\end{equation}
So we see that the scalar amplitude also Lorentz contracts, another
confirmation of our assertion that the Compton wavelength is not a
minimum width. (Note that the overall normalization, which is not
conserved under boosts for this function, also changes.)

\section{Wavepacket spreading}

Almeida and Jabs \cite{Almeida1984} treat the spreading of relativistic
wavepackets of the form of Eq. (\ref{eq:2.8}) in generality for the
spinless case. (Naumov \cite{Naumov2013} treats the spreading of
position probability amplitudes and of the scalar amplitude, but does
not consider the small $\sigma_{x}$ limit.) They consider the positive
energy to be of the form
\begin{equation}
\omega(\boldsymbol{p})=+\sqrt{\boldsymbol{p}^{2}+m^{2}},\label{eq:4.1}
\end{equation}
so that the relativistic velocity is
\begin{equation}
\boldsymbol{\beta}=\frac{\partial\omega}{\partial\boldsymbol{p}}=\frac{\boldsymbol{p}}{\omega}.\label{eq:4.2}
\end{equation}
They use a general result due to Bradford \cite{Bradford1976} that
the average position and total variance in position are given by
\begin{align}
\langle\,\boldsymbol{x}(t)\,\rangle & =\langle\,\boldsymbol{x}(0)\,\rangle+\langle\,\boldsymbol{\beta}\,\rangle t,\nonumber \\
\sigma^{2}(t) & =\sigma^{2}(0)+\{\langle\,\boldsymbol{\beta}^{2}\,\rangle-\langle\,\boldsymbol{\beta}\,\rangle^{2}\}t^{2},\label{eq:4.3}
\end{align}
respectively, where $\sigma^{2}(t)$ is the total variance
\begin{equation}
\sigma^{2}(t)=\sigma_{x}(t)^{2}+\sigma_{y}^{2}(t)+\sigma_{z}^{2}(t).\label{eq:4.4}
\end{equation}

Almeida and Jabs assume that the initial average position is the origin
($\langle\,\boldsymbol{x}(0)\,\rangle=\boldsymbol{0}$, accomplished
just with a phase change to $\Psi(p)$) and that the average velocity
vanishes ($\langle\,\boldsymbol{\beta}\,\rangle=0$, accomplished
with a translation of $\Psi(p)$ in momentum space). Then they require
that the momentum probability distribution, $|\Psi(p)|^{2}$, be such
that it becomes wider and flatter as the variance of momentum increases.
This eliminates the case of two peaks that become further apart as
the variance of momentum increases. Under only these assumptions,
they demonstrate that what they define as the spreading velocity,
\[
v_{\mathrm{Sp}}=\frac{\partial\sigma(t)}{\partial t},
\]
remains less than the speed of light but approaches it asymptotically
for large times.

We note that in three dimensions, $\sigma(t)$ is not a good measure
of the spreading rate. In the spherically symmetric case, for example,
it is $\sqrt{3}$ times any of the three individual widths, such as
$\sigma_{x}(t).$ Thus we find a stronger bound in the case of the
spherically symmetric Gaussian wavefunction (Eq. \ref{eq:13}), by
evaluating the expectation values in Eq. (\ref{eq:4.3}),
\[
\sigma_{x}(t)\rightarrow\sqrt{\sigma_{x}^{2}(0)+\frac{1}{3}t^{2}}\quad\mathrm{as}\ \sigma_{x}(0)\rightarrow0,
\]
with limiting spreading rate $\partial\sigma_{x}/\partial t\rightarrow1/\sqrt{3}$
as $t\rightarrow\infty.$

Thus this causality requirement does not place any lower limit on
the spatial extent of a wavepacket.

\section{Sub-minimal scalar amplitudes}

We have written the time-dependent position probability amplitude
as Eq. (\ref{eq:2.8}). Since $\Psi(p)$ is a momentum probability
amplitude, we require it to be square-integrable for physical states,
to make Eq.~(\ref{eq:2.3}) finite. A general positive energy solution
of the Klein-Gordon equation can be written
\begin{equation}
\varphi(x)=\int\frac{d^{3}p}{\omega}e^{-ip\cdot x}\,\Phi(p).\label{eq:5.1}
\end{equation}
We require square-integrability of $\Phi(p)/\sqrt{\omega}.$ Otherwise
$\Psi(p)$ and $\Phi(p)$ are arbitrary. The only differences between
these two expressions are the physical interpretation given to the
quantities involved and the intended Lorentz transformation properties
of the functions of position or momentum.

So we may choose
\begin{equation}
\Phi(p)=\mathcal{N}(\frac{\sigma_{p}}{m})\frac{\omega}{\sqrt{m}}\,\frac{e^{-|\boldsymbol{p}|^{2}/4\sigma_{p}^{2}}}{(2\pi\sigma_{p}^{2})^{\frac{3}{4}}},\label{eq:5.2}
\end{equation}
with $\Phi(p)/\sqrt{\omega}$ square integrable, where $\mathcal{N}(\frac{\sigma_{p}}{m})$
is a dimensionless normalization factor. This gives
\begin{equation}
\varphi(0,\boldsymbol{x})=\mathcal{N}(\frac{\sigma_{p}}{m})\frac{1}{\sqrt{m}}\frac{e^{-|\boldsymbol{x}|^{2}/4\sigma_{x}^{2}}}{(2\pi\sigma_{x}^{2})^{\frac{3}{4}}}.\label{eq:5.3}
\end{equation}
The result is a function intended to transform as a scalar function.
In this one frame at one time, its spatial width can take \textit{any}
nonzero value, $\sigma_{x}=1/2\sigma_{p},$ including values smaller
than the Compton wavelength.

\section{Conclusions}

We have provided four arguments against the claim that relativistic
quantum mechanics implies a minimum width for a particle, of order
its Compton wavelength. We find that the origin of the misconception
is the unjustified use of the scalar amplitude as a measure of localization,
instead of the relativistic probability amplitude. In relativity,
a minimum length scale is meaningless since observation from a boosted
frame will see Lorentz contraction. We reviewed the results of Almeida
and Jabs \cite{Almeida1984} on the spreading of wavepackets. They
demonstrated that a causality condition was always satisfied: the
long time spreading rate of relativistic wavepackets remained less
than the speed of light no matter how small the initial wavepacket.
Lastly, we constructed an infinite set of scalar amplitudes that violate
the lower bound on wavepacket size.

\bibliographystyle{apsrev4-1}

\end{document}